\documentstyle[sprocl,epsfig]{article}

\def\Journal#1#2#3#4{{#1} {\bf #2}, #3 (#4)}

\def\NPB{{\em Nucl. Phys.} B}
\def\PLB{{\em Phys. Lett.}  B}

\def\PRD{{\em Phys. Rev.} D}
\def\ZPC{{\em Z. Phys.} C}
\def\EPC{{\em Eur. Phys. J.} C}

\begin{document}

\title{DETERMINING THE SPIN STRUCTURE OF THE PHOTON AT FUTURE COLLIDERS}

\author{M.\ STRATMANN}

\address{Institute for Theoretical Physics, University of Regensburg,\\
D-93040 Regensburg, Germany}

\maketitle
\abstracts{It is demonstrated that measurements of the
spin asymmetry for di-jet production at future polarized colliders
appear to be particularly suited for a first determination of
the so far unmeasured parton densities of circularly polarized
photons.}

\section{$\Delta f^{\gamma}$: Framework and Present Status}
Nothing is known experimentally about the parton content of
circularly polarized photons, defined by
%
$
\Delta f^{\gamma}(x,Q^2) \equiv f_+^{\gamma_{+}}(x,Q^2) -  
f_-^{\gamma_{+}}(x,Q^2)
$,
%
where $f_+^{\gamma_{+}}$ $(f_-^{\gamma_{+}})$ 
denotes the density of a parton $f$ with helicity `+' (`$-$') 
in a photon with helicity `+',
and the next round of spin experiments, {\sc Compass} and {\sc Rhic}, is not
sensitive to these distributions either.
The $\Delta f^{\gamma}$ contain information different
from that contained in the unpolarized ones,
$f^{\gamma}$, and their measurement is vital for a {\em complete} 
understanding of the partonic structure of photons.
%
It has been demonstrated \cite{ref:svlinear} 
that measurements of the structure function $g_1^{\gamma}$
and of di-jet spin asymmetries at a future polarized linear $e^+e^-$
collider can provide valuable information about $\Delta f^{\gamma}$.
Di-jet spin asymmetries at {\sc Hera} running in a polarized collider
mode, appear to be equally promising 
\cite{ref:heraresults1,ref:heraresults2}. Here we will focus on two other 
recent proposals for a polarized $ep$ collider: {\sc eRhic} and {\sc THera}.
As in \cite{ref:svlinear,ref:heraresults1,ref:heraresults2} we will exploit the 
predictions of two very different models for the $\Delta f^{\gamma}$ 
\cite{ref:gsv}, and study the sensitivity of di-jet production to these 
unknown quantities. In the first case (`maximal scenario') 
we saturate the positivity 
bound $|\Delta f^{\gamma}(x,Q^2)| \le f^{\gamma}(x,Q^2)$
at a low input scale $\mu\simeq 0.6\,{\rm{GeV}}$, using the
unpolarized GRV densities $f^{\gamma}$ \cite{ref:grvphot}.
The other extreme input (`minimal scenario') is defined by
a vanishing hadronic input at the scale $\mu$.
We limit ourselves to leading order (LO) QCD, 
which is entirely sufficient for our purposes; however
both scenarios can be straightforwardly extended to 
next-to-leading order \cite{ref:nloletter}.

\section{$\Delta f^{\gamma}$: Tests and Signatures}
The generic expression for polarized {\em resolved} photoproduction of two jets 
with laboratory system rapidities $\eta_1$, $\eta_2$ 
and transverse momentum $p_T$
reads in LO
\begin{equation} 
\label{eq:wq2jet}
\frac{d^3 \Delta \sigma}{dp_T d\eta_1 d\eta_2} = 2 p_T
\sum_{f^e,f^p} x_e \Delta f^e (x_e,\mu_f^2) x_p \Delta f^p (x_p,\mu_f^2)
\frac{d\Delta \hat{\sigma}}{d\hat{t}}\; ,
\end{equation}
where 
$x_e \equiv p_T/(2 E_e) \left( e^{-\eta_1} + e^{-\eta_2} \right)$ and
$x_p \equiv p_T/(2 E_p) \left( e^{\eta_1} + e^{\eta_2} \right)$.
The $\Delta f^p$ and $\Delta f^e$
in (\ref{eq:wq2jet}) denote the spin-dependent parton 
densities of the proton and electron, i.e., photon, respectively,
see \cite{ref:heraresults2}.
The key feature of {\em di}-jet production is that a measurement of both
jet rapidities allows for fully reconstructing the kinematics of the 
underlying hard subprocess and thus for determining $x_{\gamma}=x_e/y$
experimentally, with $y$ being the fraction of the electron's energy taken by 
the photon. In this way it becomes possible to 
suppress the {\em direct} ($x_{\gamma}=1$) contribution by, e.g.,
scanning different bins in $x_{\gamma}$, cf.\ \cite{ref:h1}.

Fig.\ 1 shows the di-jet spin asymmetries 
$A^{\mathrm{2-jet}}\equiv d\Delta\sigma/d\sigma$ at
{\sc eRhic} ($\sqrt{s}=100\;\mathrm{GeV}$) 
and {\sc THera} ($\sqrt{s}=950\;\mathrm{GeV}$) for three
bins in $x_{\gamma}$, using $\mu_f=p_T$, $0.2\le y\le0.85$ and
similar cuts for $|\eta_1 - \eta_2|$ and $(\eta_1+\eta_2)/2$ as in \cite{ref:h1}.
\begin{figure}[bh]
\begin{center}
\vspace*{-0.8cm}
\mbox{\epsfig{figure=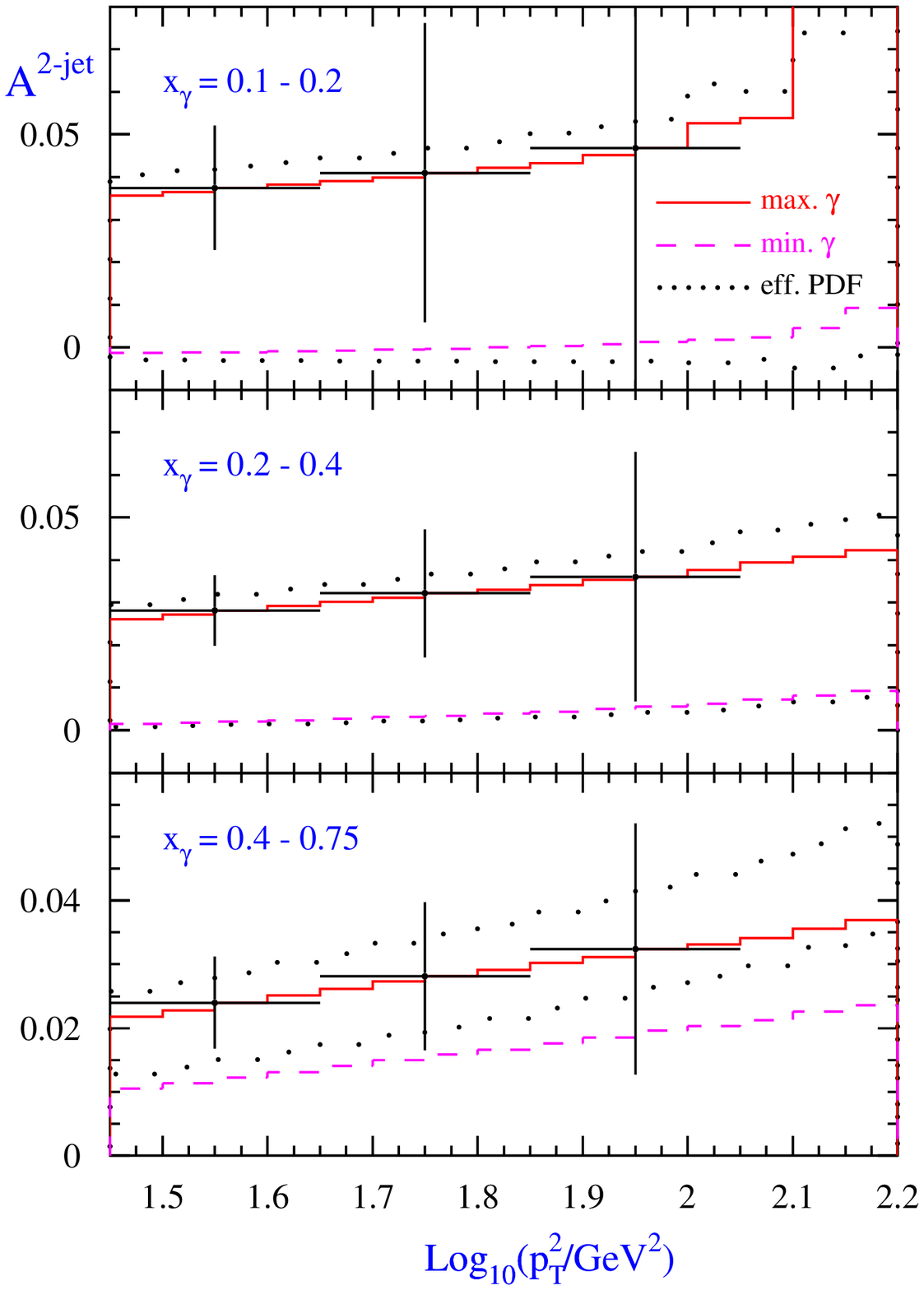,width=0.48\textwidth,clip=}}
\mbox{\epsfig{figure=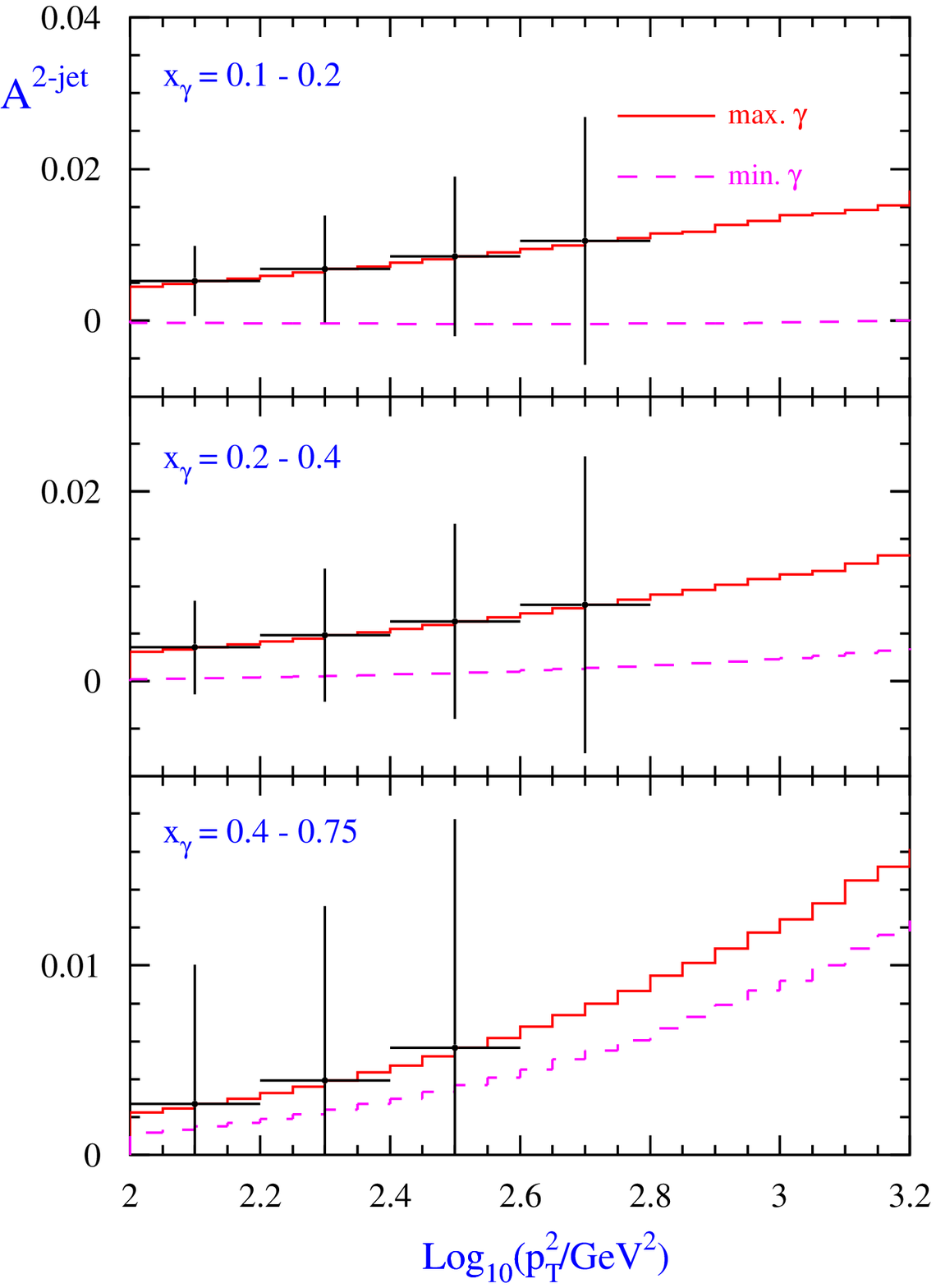,width=0.48\textwidth,clip=}}
\end{center}
\vspace*{-0.3cm}
\caption{Predictions for $A^{\mathrm{2-jet}}$ in LO
(left: {\sc eRhic}, right: {\sc THera}) using the two scenarios for
$\Delta f^{\gamma}$ as described above and the
GRSV $\Delta f^p$ densities$^8$.
Also shown are the expected statistical errors for such a measurement
assuming 70\% beam polarizations and ${\cal{L}}=200\;\mathrm{pb}^{-1}$.}
\end{figure}

To actually unfold information on $\Delta f^{\gamma}$ 
it is useful to introduce the concept of `effective parton
densities' \cite{ref:effpdf}. Although  $A^{\mathrm{2-jet}}$ 
is dominated by $gg$ scattering, {\em all}
QCD subprocesses contribute.
In the unpolarized case it was shown \cite{ref:effpdf}
that the ratios of dominant subprocesses
are roughly constant, i.e., $qq'/qg\simeq qg/gg\simeq 4/9$, such that
the jet cross section factorizes approximately 
into some effective parton densities times a {\em single} subprocess 
cross section.
In the polarized case this factorization is slightly broken as
$qq'/qg \neq qg/gg$. However, the approximation still
works surprisingly well at a level of $5-10\%$ accuracy, and 
the appropriate effective densities are given by \cite{ref:heraresults2}
(see also ref.\ \cite{ref:effpdfpol})
\begin{equation}
\label{eq:effpdf}
\Delta f_{\mathrm{eff}}^{\gamma} = \sum_q (\Delta q^{\gamma}
                                   +\Delta \bar{q}^{\gamma})+
                                   \frac{11}{4} \Delta g^{\gamma}
\end{equation} 
such that the polarized double resolved jet cross section can be
expressed as
\begin{equation}
\label{eq:effxsec}
\Delta \sigma^{\mathrm{2-jet}} \simeq \Delta f_{\mathrm{eff}}^{\gamma} \otimes
                     \Delta f_{\mathrm{eff}}^{p} \otimes
                     \Delta \hat{\sigma}_{qq'\rightarrow qq'}\,\,.
\end{equation}
As can be inferred from the l.h.s.\ of Fig.~1, the effective parton density 
approximation (dotted lines) works very well indeed. 
It is only for large $p_T$ that the deviations from the exact results 
become more pronounced.  

Given the error bars shown in Fig.\ 1, the prospects for 
distinguishing between different scenarios for $\Delta f^{\gamma}_{\mathrm{eff}}$ 
are rather promising for {\sc eRhic} (but remote for {\sc THera}
where only luminosities of ${\cal{O}}(10\;\mathrm{pb}^{-1})$ seem to be
realistic) {\em provided} the $\Delta f^{p}_{\mathrm{eff}}$, also entering 
(\ref{eq:effxsec}), are known fairly well, which is clearly not the case yet. 
However, our ignorance of the $\Delta f^{p}$ 
will be vastly reduced by the upcoming polarized $pp$ collider 
{\sc Rhic} and ongoing efforts in the fixed target sector.
It should be kept in mind that so far {\em nothing at all} is known about
the $\Delta f^{\gamma}$, and even to establish the very existence
of a resolved component also in the spin-dependent case would be
an important step forward. 

\section*{Acknowledgments}
I am grateful to W.\ Vogelsang for a pleasant collaboration.

\section*{References}

%

\begin{thebibliography}{99}
%
\bibitem{ref:svlinear} M.\ Stratmann and W.\ Vogelsang,
{\em Nucl. Phys. B (Proc. Suppl.)} {\bf 82}, 400 (2000).
%
\bibitem{ref:heraresults1} M.\ Stratmann and W.\ Vogelsang, 
\Journal{\ZPC}{74}{641}{1997}; 
in {\em Future Physics at HERA}, G.\ Ingelman {\it et al.} (eds.), p.\ 815;
J.M.\ Butterworth {\it et al.}, in DESY-PROC-1998-01, p.~120
({\em hep-ph/9711250}).
%
\bibitem{ref:heraresults2} M.\ Stratmann, and
W.\ Vogelsang, in  DESY-PROC-1999-03, p.\ 324.
%
\bibitem{ref:gsv} M.\ Gl\"{u}ck and W.\ Vogelsang, 
\Journal{\ZPC}{55}{353}{1992}; {\em ibid.} C {\bf 57}, 309 (1993);
M.\ Gl\"{u}ck {\it et al.}, \Journal{\PLB}{337}{373}{1994}.
%
\bibitem{ref:grvphot} M.\ Gl\"{u}ck, E.\ Reya, and A.\ Vogt, 
\Journal{\PRD}{46}{1973}{1992}.
%
\bibitem{ref:nloletter} M.\ Stratmann and W.\ Vogelsang, 
\Journal{\PLB}{386}{370}{1996}.
%
\bibitem{ref:h1} H1 collab.: C.\ Adloff {\it et al.}, 
\Journal{\EPC}{1}{97}{1998}.
%
\bibitem{ref:grsv} M.\ Gl\"{u}ck {\it et al.}, \Journal{\PRD}{53}{4775}{1996}.
%
\bibitem{ref:effpdf} B.L.\ Combridge and C.J.\ Maxwell, 
\Journal{\NPB}{239}{429}{1984}.
%
\bibitem{ref:effpdfpol} D.\ Indumathi {\it et al.},
\Journal{\ZPC}{56}{427}{1992}.
%
\end{thebibliography}
\end{document}